\documentstyle[twoside,epsf]{article}

\catcode`\@=11
\long\def\@makefntext#1{
\protect\noindent \hbox to 3.2pt {\hskip-.9pt  
$^{{\eightrm\@thefnmark}}$\hfil}#1\hfill}		

\def\@makefnmark{\hbox to 0pt{$^{\@thefnmark}$\hss}}	
	
\def\ps@myheadings{\let\@mkboth\@gobbletwo
\def\@oddhead{\hbox{}
\rightmark\hfil\eightrm\thepage}   
\def\@oddfoot{}\def\@evenhead{\eightrm\thepage\hfil
\leftmark\hbox{}}\def\@evenfoot{}
\def\sectionmark##1{}\def\subsectionmark##1{}}

\oddsidemargin=\evensidemargin
\newcounter{sectionc}\newcounter{subsectionc}\newcounter{subsubsectionc}
\renewcommand{\section}[1] {\vspace{12pt}\addtocounter{sectionc}{1} 
\setcounter{subsectionc}{0}\setcounter{subsubsectionc}{0}\noindent 
	{\tenbf\thesectionc. #1}\par\vspace{5pt}}
\renewcommand{\subsection}[1] {\vspace{12pt}\addtocounter{subsectionc}{1} 
	\setcounter{subsubsectionc}{0}\noindent 
	{\bf\thesectionc.\thesubsectionc. {\kern1pt \bfit #1}}\par\vspace{5pt}}
\renewcommand{\subsubsection}[1] {\vspace{12pt}\addtocounter{subsubsectionc}{1}
	\noindent{\tenrm\thesectionc.\thesubsectionc.\thesubsubsectionc.
	{\kern1pt \tenit #1}}\par\vspace{5pt}}
\newcommand{\nonumsection}[1] {\vspace{12pt}\noindent{\tenbf #1}
	\par\vspace{5pt}}

\newcounter{appendixc}
\newcounter{subappendixc}[appendixc]
\newcounter{subsubappendixc}[subappendixc]
\renewcommand{\thesubappendixc}{\Alph{appendixc}.\arabic{subappendixc}}
\renewcommand{\thesubsubappendixc}
	{\Alph{appendixc}.\arabic{subappendixc}.\arabic{subsubappendixc}}

\renewcommand{\appendix}[1] {\vspace{12pt}
        \refstepcounter{appendixc}
        \setcounter{figure}{0}
        \setcounter{table}{0}
        \setcounter{lemma}{0}
        \setcounter{theorem}{0}
        \setcounter{corollary}{0}
        \setcounter{definition}{0}
        \setcounter{equation}{0}
        \renewcommand{\thefigure}{\Alph{appendixc}.\arabic{figure}}
        \renewcommand{\thetable}{\Alph{appendixc}.\arabic{table}}
        \renewcommand{\theappendixc}{\Alph{appendixc}}
        \renewcommand{\thelemma}{\Alph{appendixc}.\arabic{lemma}}
        \renewcommand{\thetheorem}{\Alph{appendixc}.\arabic{theorem}}
        \renewcommand{\thedefinition}{\Alph{appendixc}.\arabic{definition}}
        \renewcommand{\thecorollary}{\Alph{appendixc}.\arabic{corollary}}
        \renewcommand{\theequation}{\Alph{appendixc}.\arabic{equation}}
        \noindent{\tenbf Appendix \theappendixc #1}\par\vspace{5pt}}
\newcommand{\subappendix}[1] {\vspace{12pt}
        \refstepcounter{subappendixc}
        \noindent{\bf Appendix \thesubappendixc. {\kern1pt \bfit #1}}
	\par\vspace{5pt}}
\newcommand{\subsubappendix}[1] {\vspace{12pt}
        \refstepcounter{subsubappendixc}
        \noindent{\rm Appendix \thesubsubappendixc. {\kern1pt \tenit #1}}
	\par\vspace{5pt}}

\newcommand{\textlineskip}{\baselineskip=13pt}
\newcommand{\smalllineskip}{\baselineskip=10pt}

\def\eightcirc{
\begin{picture}(0,0)
\put(4.4,1.8){\circle{6.5}}
\end{picture}}
\def\eightcopyright{\eightcirc\kern2.7pt\hbox{\eightrm c}} 

\newcommand{\copyrightheading}[1]
	{\vspace*{-2.5cm}\smalllineskip{\flushleft
	{\footnotesize International Journal of Modern Physics C, #1}\\
	{\footnotesize $\eightcopyright$\, World Scientific Publishing
	 Company}\\
	 }}

\newcommand{\publisher}[2]{{\begin{center}\footnotesize\smalllineskip 
	Received #1\\
	Revised #2
	\end{center}
	}}

\def\abstracts#1#2#3{{
        \centering{\begin{minipage}{4.5in}\baselineskip=10pt\footnotesize
        \parindent=0pt #1\par 
        \parindent=15pt #2\par
        \parindent=15pt #3
        \end{minipage}}\par}} 

\def\keywords#1{{
	\centering{\begin{minipage}{4.5in}\baselineskip=10pt\footnotesize
	{\footnotesize\it Keywords}\/: #1
	\end{minipage}}\par}}

\renewenvironment{thebibliography}[1]
        {\frenchspacing
	 \ninerm\baselineskip=11pt
         \begin{list}{\arabic{enumi}.}
        {\usecounter{enumi}\setlength{\parsep}{0pt}     
	 \setlength{\leftmargin 12.7pt}{\rightmargin 0pt} 
         \setlength{\itemsep}{0pt} \settowidth
	{\labelwidth}{#1.}\sloppy}}{\end{list}}

\newcounter{itemlistc}
\newcounter{romanlistc}
\newcounter{alphlistc}
\newcounter{arabiclistc}

\newcommand{\fcaption}[1]{
        \refstepcounter{figure}
        \setbox\@tempboxa = \hbox{\footnotesize Fig.~\thefigure. #1}
        \ifdim \wd\@tempboxa > 5in
           {\begin{center}
        \parbox{5in}{\footnotesize\smalllineskip Fig.~\thefigure. #1}
            \end{center}}
        \else
             {\begin{center}
             {\footnotesize Fig.~\thefigure. #1}
              \end{center}}
        \fi}

\newcommand{\tcaption}[1]{
        \refstepcounter{table}
        \setbox\@tempboxa = \hbox{\footnotesize Table~\thetable. #1}
        \ifdim \wd\@tempboxa > 5in
           {\begin{center}
        \parbox{5in}{\footnotesize\smalllineskip Table~\thetable. #1}
            \end{center}}
        \else
             {\begin{center}
             {\footnotesize Table~\thetable. #1}
              \end{center}}
        \fi}

\def\@citex[#1]#2{\if@filesw\immediate\write\@auxout
	{\string\citation{#2}}\fi
\def\@citea{}\@cite{\@for\@citeb:=#2\do
	{\@citea\def\@citea{,}\@ifundefined
	{b@\@citeb}{{\bf ?}\@warning
	{Citation `\@citeb' on page \thepage \space undefined}}
	{\csname b@\@citeb\endcsname}}}{#1}}

\newif\if@cghi
\def\cite{\@cghitrue\@ifnextchar [{\@tempswatrue
	\@citex}{\@tempswafalse\@citex[]}}
\def\citelow{\@cghifalse\@ifnextchar [{\@tempswatrue
	\@citex}{\@tempswafalse\@citex[]}}
\def\@cite#1#2{{$\null^{#1}$\if@tempswa\typeout
	{IJCGA warning: optional citation argument 
	ignored: `#2'} \fi}}

\def\pmb#1{\setbox0=\hbox{#1}
	\kern-.025em\copy0\kern-\wd0
	\kern.05em\copy0\kern-\wd0
	\kern-.025em\raise.0433em\box0}

\def\fnt#1#2{\footnotetext{\kern-.3em
	{$^{\mbox{\scriptsize #1}}$}{#2}}}

\def\fpage#1{\begingroup
\voffset=.3in
\thispagestyle{empty}\begin{table}[b]\centerline{\footnotesize #1}
	\end{table}\endgroup}

\def\runninghead#1#2{\pagestyle{myheadings}
\markboth{{\protect\footnotesize\it{\quad #1}}\hfill}
{\hfill{\protect\footnotesize\it{#2\quad}}}}
\font\tenrm=cmr10
\font\tenit=cmti10 
\font\tenbf=cmbx10
\font\bfit=cmbxti10 at 10pt
\font\ninerm=cmr9

\font\eightrm=cmr8

\def\qed{\hbox{${\vcenter{\vbox{			
   \hrule height 0.4pt\hbox{\vrule width 0.4pt height 6pt
   \kern5pt\vrule width 0.4pt}\hrule height 0.4pt}}}$}}

\def\bsc{{\sc a\kern-6.4pt\sc a\kern-6.4pt\sc a}}  	
\def\bflatex{\bf L\kern-.30em\raise.3ex\hbox{\bsc}\kern-.14em 
T\kern-.1667em\lower.7ex\hbox{E}\kern-.125em X} 

\begin{document}

\runninghead{M. Isobe} {Algorithm for Molecular Dynamics Simulation in Hard Disk System}

\normalsize\textlineskip
\thispagestyle{empty}
\setcounter{page}{1}

\copyrightheading{}			

\vspace*{0.88truein}

\fpage{1}
\centerline{\bf SIMPLE AND EFFICIENT ALGORITHM FOR LARGE}
\vspace*{0.035truein}
\centerline{\bf SCALE MOLECULAR DYNAMICS SIMULATION}
\vspace*{0.035truein}
\centerline{\bf IN HARD DISK SYSTEM}
\vspace*{0.37truein}
\centerline{\footnotesize MASAHARU ISOBE}
\vspace*{0.015truein}
\centerline{\footnotesize\it Department of Physics, Kyushu University 33, Fukuoka 812-8581, Japan}
\baselineskip=10pt
\centerline{\footnotesize\it E-mail: isobe@stat.phys.kyushu-u.ac.jp}
\vspace*{0.225truein}
\publisher{25 June 1999}{21 July 1999}

\vspace*{0.21truein}
\abstracts{
A simple and efficient algorithm of the molecular-dynamics simulation of 
the hard disk system based on the Event-Driven method is developed. From the 
analysis of algorithm, the complexity is ${\cal O}(\log N)$ per 1 event, and 
the constant coefficient of the complexity is smaller than conventional 
efficient algorithm based on the concept of Cell-Crossing Event. The maximum 
performance of more than 460 millions of collisions per CPU-hour on the 
Alpha600 compatible in a 2500 particle system is achieved. An extension to 
the infinite-space system based on this algorithm is also proposed.
}{}{}

\vspace*{10pt}


\vspace*{1pt}\textlineskip	

\section{Introduction}
\label{sec:1-0}
\noindent
The molecular dynamics method (MD) was worked out in a paper for the first 
time by Alder and Wainwright in 1957,\cite{alder_1957} and they subsequently
 wrote a series of works.\cite{alder_1959,alder_1960,alder_1962} Their 
simulations were performed with many hard disks (or hard spheres in 3D) 
near the liquid-solid phase-transition point, and they found that the system 
was crystallized despite the particle had only repulsive force. These discoveries 
overturned the common opinion of those days, and greatly influenced the 
development of the study in computer simulations.
 
In the hard disk system, the dynamics consists of only collisions and 
straight-line movement. Since distinct events occur one after another in time, 
we do not need to integrate the differential equations with a constant time 
step based on Newton's equation of motion. The method that is based on the 
finite constant time step and integration with the equations of particles step 
by step in time is sometimes called ``Time-Step-Driven Molecular Dynamics''
(TSDMD). On the other hand, in the hard disk system, the simulation that 
proceeds based on events is called ``Event-Driven Molecular Dynamics'' (EDMD).
Compared with TSDMD simulation, the algorithm of EDMD simulation is completely 
different. We need the knowledge of an advanced algorithm and a data structure 
to perform the efficient simulation in EDMD. The strategy of direct computation 
of particle-pairs result in the complexity ${\cal O}(N^2)$ for large particle 
number $N$. The point of improvement in the speed in hard disk system is how 
well we deal with the queue of a future event and the data structure.
 
The improvement of complexity in the algorithm of large-scale hard disk system 
was developed by Rapaport.\cite{rapaport_1980,rapaport_1995} This 
algorithm is based on the concept of sub-cell method,\cite{erpenbeck_1977}
and both Collision Event and Cell-Crossing Event are stored into 
Multiple-Event Time List. Then the minimum time is searched by Binary Search 
Tree(BST).\cite{knuth_1973} When the event occurs, the particle-pair or the 
particle --- sub-cell respectively relevant to Collision Event or Cell-Crossing 
Event is deleted, and collision time for the particle relevant to the event is 
re-computed and new nodes are created. The BST is reconstructed by inserting 
these new nodes, and the minimum time is searched. On this algorithm, the 
averaged complexity per event becomes ${\cal O}(\log N)$, and the reduction 
of a large amount of computation is realized.

However, since the algorithm of Rapaport is very complicated and difficult to 
understand, several algorithms to simplify a data structure and 
improve the efficiency in the large-scale molecular dynamics simulation were 
proposed in the 90s.\cite{lubachevsky_1991,shida_1992,krantz_1996}
Mar\'{\i}n et al.\cite{marin_1993} developed the technique of 
Local Minima Algorithm (LMA) to avoid additional re-computation for Event List.
When we actually schedule future event list, LMA put only the minimum event 
time relevant to each particle into Complete Binary Tree (CBT). In 1995, 
Mar\'{\i}n and Cordero\cite{marin_1995} compared various 
${\cal O}(\log N)$ searching algorithms actually in EDMD simulation, 
systematically. They concluded that the efficiency of Complete Binary Tree 
(CBT) was the most suitable for hard disks system in all density regions. 
If the number of particles increases, for CBT it was clearly shown that 
efficiency increased significantly when compared with other searching algorithms.
If one compare CBT with BST, their complexity are the same, 
i.e., ${\cal O}(\log N)$; but the algorithm of CBT is much simpler than 
that of BST, consequently CBT requires less memory space and realizes better 
actual performance. 

In this paper, we developed an algorithm based on a strategy different from 
that of Cell-Crossing type.
The algorithm is extended to Exclusive Particle Grid 
Method (EPGM) developed by Form et al.,\cite{form_1993} (Sec. 2). Then, a 
bookkeeping method\cite{verlet_1967} is applied (Sec. 3). Compared with the 
Cell-Crossing type, our algorithm extended the concept of Linked-Cell 
Method\cite{quentrec_1975,hockney_1981} and Neighbor List, which are often 
used in TSDMD to carry out an efficient simulation.\cite{allen_1987}
From the analysis of complexity, we show our algorithm is smaller than the 
complexity of Cell-Crossing type. By an empirical evaluation of the simulation 
in hard disk systems, our code could be shown to perform better than that of 
any previously-published works.

The infinite volume extension of the sub-cell method is usually considered
impossible because the required memory is proportional to the volume in
the conventional sub-cell method.\cite{shida_1992}
We developed the method of compressing information about the infinite sub-cells 
into limited finite arrays. In addition, the hashing method, which is considered 
the most efficient searching algorithm, is applied to our method in order to 
pull out the information on a neighbor cell in high speed. It is 
found that the hashing method can be applied especially easily on our method.

Various applications in a very wide-ranging field will be possible by changing 
the external field and the collision rule in the large-scale EDMD. Typical 
examples performed by EDMD so far are: phase transition in the 
equilibrium system (solid-liquid transition and two-dimensional melting),
\cite{alder_1957,alder_1959,alder_1960,alder_1962}
the nonequilibrium fluid system (e.g., Rayleigh-B\'{e}nard convection),
\cite{mareschal_1988,rapaport_1988} the nonequilibrium chemical-reaction
system,\cite{kawakatu_1987} the nonequilibrium dissipative system 
(granular system),\cite{goldhirsch_1993,mcnamara_1996,bizon_1998}
random disk packing.\cite{lubachevsky_1990,lubachevsky2_1991}
When large-scale computation becomes possible in these systems, 
the value of simulation with a hard disk system must increase significantly.

The outline of this paper is as follows: in Secs. 2 and 3, the algorithms 
developed are explained. The comparisons with the algorithm of 
Cell-Crossing type by analyzing the complexities are shown in Sec. 4. The 
empirical evaluation is given in Sec. 5. The extension to the infinite 
system based on developed algorithms is explained in Sec. 6. In Sec. 7, a 
short summary and a final comment are presented.

\section{Extended Exclusive Particle Grid Method}
\label{sec:2-0}
\noindent
The sub-cell method is often used to achieve the significant reduction of 
computation. In the EDMD, the complexity of producing and updating future 
event list for restoring event time of each particle-pair are reduced to 
${\cal O}(N)$ and ${\cal O}(1)$, respectively, by the sub-cell method. 
LCM is the method of dividing the system into small sub-cells. When the size of 
the sub-cell is bigger than a particle diameter, we will have a difficulty 
in coding a program, because we do not know how many particles enter in each 
sub-cell.
Therefore, link lists must be prepared for the particles in each sub-cell. 
 
On the other hand, another efficient sub-cell method, called Exclusive 
Particle Grid Method (EPGM), was independently developed by Buchholtz and 
P\"{o}schel \cite{buchholtz_1993} and Form, Ito, and 
Kohring \cite{form_1993} to simulate the soft-core granular particle 
system in TSDMD. In this method, there is only one particle in each sub-cell.
Though the EPGM is essentially the method of putting a particle to one 
sub-cell in LCM, it does not need to use pointers for neighbor sub-cells or 
link list. Here sub-cells in EPGM are called ``grid''. In the EPGM, the length 
of grid $l_{gx}$ is determined by the following inequality:
 
\begin{equation}
\sigma < l_{gx} < \sqrt{2}\sigma,
\end{equation}
 
\noindent
where $\sigma$ is the radius of particle.
One example of the number of grids $n_{gx}, n_{gy}$ in the system of length 
$l_x, l_y$ and the length of grid $l_{gx}, l_{gy}$ can be calculated by the 
following equations.

\begin{equation}
      l_{gx}=INT(l_x/(\sqrt{2}\sigma))+1
\end{equation}
\begin{equation}
      l_{gy}=INT(l_y/(\sqrt{2}\sigma))+1
\end{equation}
\begin{equation}
      n_{gx}=l_x/l_{gx}
\end{equation}
\begin{equation}
      n_{gy}=l_y/l_{gy}.
\end{equation}

\begin{figure}
\begin{center}
\leavevmode
\epsfxsize=8.5cm
\epsfbox{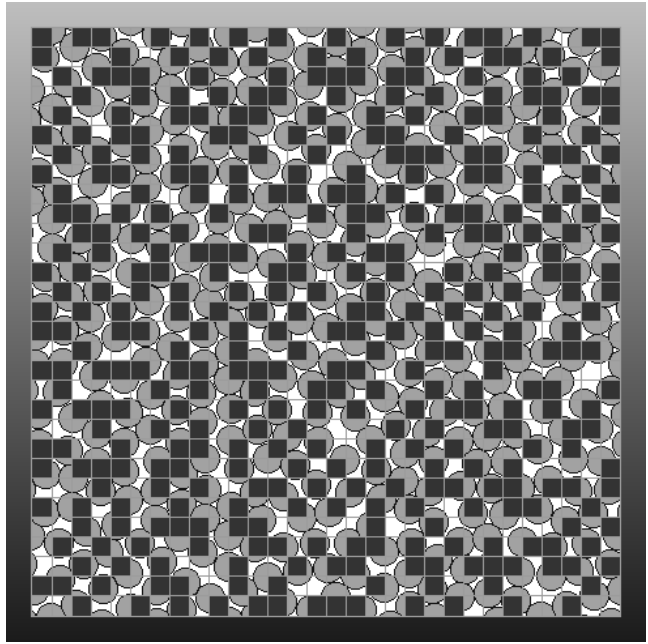}
\end{center}
\caption{A typical example of the mapping pattern using EPGM when packing 
fraction $\nu=0.70$. Both the position of hard disks and the occupied 
lattices are shown.}
\label{fig:2-1}
\end{figure}

\noindent
Note that the total number of grids is $n_g=n_{gx}\times n_{gy}$.
As an analogy to the lattice spin system, EPGM is regarded as $N+1$-states 
Potts model of the square lattice system.
This is because the particles are completely mapped into each lattice 
(i.e., one grid corresponds to one particle respectively), and we put 0 into 
the rest of grids in which there is no particle (Fig.~\ref{fig:2-1}).
Since continuous and random positions of the particles are mapped into the 
lattice, the specification of neighbor particles becomes very easy. 

\noindent
Form et al.\cite{form_1993} applied this algorithm in the high-density 
soft-core granular system with short-distance interaction in TSDMD.
They achieved high efficiency on a vector computer. Based on this algorithm, 
the extension of EPGM to EDMD in the hard disk system is developed.
 
When the system is in high-density, EPGM can be simply applied to EDMD.
For a candidate of the next colliding particle-pairs, we need to search only 24 
neighbor grids, which form the square mask. If neighbor grid is not 0, the 
collision times of candidates of colliding particle-pairs are computed only 
by the registered particle number in the square mask. We call these 24 neighbor grids 
MIN, because this is the minimum mask in the simulation of EDMD. Note that if 
the smaller mask is used, the computation will break down during the 
simulation, since a possibility of overlap between a central particle 
in the mask and particles outside the mask occurs. When EPGM is applied in 
the high-density system, the computation is optimized because a sufficient 
number of particles contained in the mask MIN. These are enough to be candidates 
of particle-pairs of collision, and only the required particles are registered 
in the mask MIN; the efficiency increases as a result of the computational 
reduction of collision time for neighbor particle-pairs.

\begin{figure}
\begin{center}
\leavevmode
\epsfxsize=8.5cm
\epsfbox{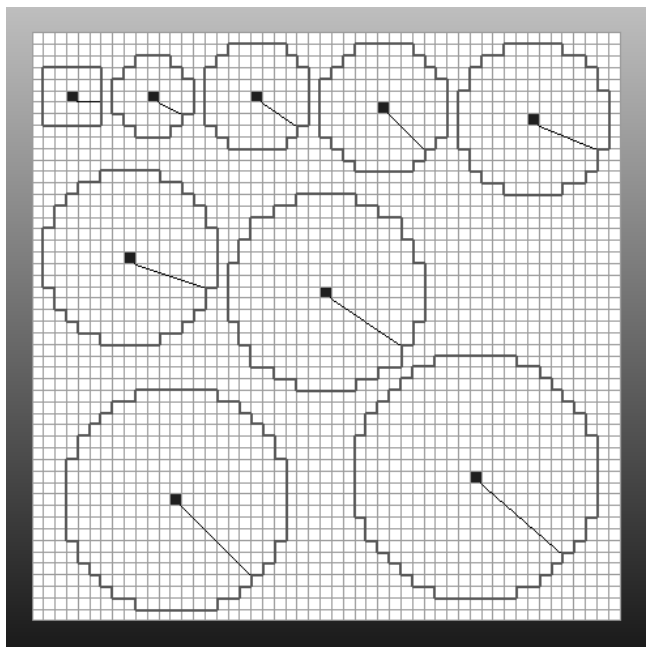}
\end{center}
\caption{Minimum mask (MIN) and larger masks, which are generated by the 
algorithm of making circle on the computer display, are shown. The solid lines 
in each mask denote the minimum distance from the central lattice of mask to 
the frame of mask.}
\label{fig:2-2}
\end{figure}

On the other hand, when the system is of low density, no sufficient number of 
particles as candidates of collision are registered in the mask MIN. 
Under such a situation, the computation will break down, since collision time 
between the central particle in the mask and particles out of the mask will 
become the next minimum time. In order to prevent this breakdown, the 
extension of EPGM is developed. The region must be extended to look for 
candidates of collision particle-pairs bigger than MIN. Since the rigorous 
isotropy of neighbors in EDMD is not necessarily demanded, the shape of the 
mask might be approximated by a rough circle. It is found that the mask 
approximated by the lattice-like grid with the circle can use the algorithm 
describing the circle on the discrete space of computer display. Figure \ref{fig:2-2}
displayed the circles from R= 3 to R= 10 on the discrete space of computer 
display, which MIN is also showed over Fig.~\ref{fig:2-2}. The total number of neighbor 
grids (mask) are 24 (MIN), 36 (R=3), 68 (R=4), 96 (R=5), 136 (R=6), 176 (R=7),
 224 (R=8), 292 (R=9), 348 (R=10), respectively. This extension of EPGM is 
called Extended Exclusive Particle Grid Method (EEPGM).

Compared with LCM, EEPGM is simple, because rough neighbor particles can be 
simply regulated by grids so that only the necessary minimum may be taken.
In EEPGM, since link list and pointers for neighbor sub-cell is not necessary, 
the required memory is also sharply reduced, the number of operations for 
setting EEPGM is small, and the program becomes very simple. Moreover, the 
extension to the infinite volume system is easy when using a hashing method 
to EEPGM. This is explained in detail in Sec. 6.
 
\section{Neighbor List and Dynamical Upper Time Cut-Off}
\label{sec:3-0}
\noindent
Though EEPGM produces a significant reduction of computation compared with 
considering all particle-pairs, it is inadequate when large number of particle 
simulation are performed. In this section, the next step in the strategy 
of the increase in efficiency based on EEPGM is developed. Here, the concept of 
Neighbor List (NL)\cite{verlet_1967} is adopted. Since the grids correspond to 
each particle, we can regard the registration of Neighbor List as already 
being completed. Therefore, we need only to search neighbor particles along in
the form of a mask.
 
In the usual way of LCM+NL, after the system is divided into sub-cells, particles 
are listed into the link list. Then neighbor particles within radius $r_{NL}$ are 
entered into Neighbor List from the link list. However, since the length of both 
lists is unknown, they must be estimated by trial and error. Although 
registration of NL is completed by one operation in EEPGM, LCM+NL must use 
two different unknown size lists, which is accompanied by a complicated 
procedure and requires difficult programming. Therefore, EEPGM (+NL) is simple, 
which means that both debugging and extension do not require excessive effort;
moreover, only the minimum nearest particles can be seen, because the system is 
divided into the minimum sub-cells, i.e., the grid. Since registration of NL is 
completed by one operation in EEPGM, efficiency is better than LCM+NL 
at a result.
 
The next improvement in speed is that the computation of collision time only 
from particle-pairs which are registered in the mask of EEPGM during the 
time $t_{NL}$. Then the complexity of EEPGM for every event is reducible to 
${\cal O}(1)$ instead of ${\cal O} (N)$. After time $t_{NL}$, the 
grids are again re-mapped in order to update neighbor particles. The time of 
Neighbor List $t_{NL}$ must be determined such that the central particle does 
not collide with a particle completely out of the mask. 
If $t_{NL}$ is too long and such a event happens, the program picks up 
a wrong pairs of particles to collide, which could result in negative 
collision time.
This conventional determination of $t_{NL}$ needs a lot of trial and error. 
In order to overcome these difficulties, $t_{NL}$ is determined by the 
following procedure. First, after completing EEPGM, the maximum velocity 
$v_{max}$ in the system is searched, and the value of its velocity is 
restored (the complexity of this searching is the same order of EEPGM 
${\cal O}(N)$). Next, time $t_{NLmin}$ of the system is 
calculated. In this calculation we suppose both the central particle and the 
particle out of the mask have the maximum velocity and those particles undergo 
head-on collision. If $t_{NL}=t_{NLmin}$, a counting mistake of collision 
pairs in the system never occurred during the time $t_{NLmin}$. The minimum 
NL distance $r_{NLmin}$ is required when $t_{NLmin}$ is calculated, which 
becomes clear from the geometry of the adopted mask shown in Fig.~\ref{fig:2-2}.
Therefore, $t_{NLmin}$ is given by 

\begin{equation}
t_{NLmin}=\frac{r_{NLmin}-2\sigma_{max}}{2v_{max}}
\end{equation}

\noindent
where $\sigma_{max}$ is the maximum radius of particle in the system.
The minimum distances $r_{min}$ for each mask are shown in Table \ref{tbl:3-1}. 

\begin{table}
\begin{center}
\begin{tabular}{|c|c|c|c|c|c|c|c|c|c|}
\hline
{} & MIN. & R=3 & R=4 & R=5 & R=6 & R=7 & R=8 & R=9 & R=10 \\
\hline
\hline
$r_{min} / l_{gx}$ & $2$ & $\sqrt{5}$ & $\sqrt{13}$ & $\sqrt{18}$ & 
$\sqrt{29}$ & $\sqrt{40}$ & $\sqrt{52}$ & $\sqrt{72}$ & $\sqrt{85}$ \\
\hline
\end{tabular}
\caption{The minimum distance for each mask.}
\label{tbl:3-1}
\end{center}
\end{table}

When we simulate in the equilibrium system, this strategy will work well 
because $t_{NLmin}$ hardly changes. However, in the nonequilibrium system 
(e.g., the relaxing process or the system with heat bath) it breaks down 
because the maximum velocity changes drastically at every step. To overcome 
this difficulty, we must check the maximum velocity for each event with energy 
increase. Fortunately the complexity of this checking process is ${\cal O}(1)$.
Therefore, in the simulation of the nonequilibrium system $t_{NL}$ will 
change one after another. We call this change in $t_{NL}$ techniques Dynamical 
Upper Time Cut-off (DUTC).
The development of DUTC, EEPGM became applicable in the nonequilibrium system.

Although we do not need to update the grid pattern for a long time 
in the high-density system, we should often do so
when we use the mask MIN in low-density systems because the grid pattern 
changes drastically.
In order to reduce the frequency of updating grid pattern, we can only use a 
bigger mask.

\section{Analysis of Complexity}
\label{sec:4-0}
\noindent
Analysis of complexity is one of important factor in estimating the efficiency 
of algorithms. In this section, a comparison of analysis of complexity between 
the algorithm of the Cell-Crossing type and the EEPGM + DUTC is shown.
The difference between the algorithm of Cell-Crossing type and the strategy of 
EEPGM + DUTC is Cell-Crossing Event itself. Therefore, especially with regard 
this point, both complexities with a constant coefficient $k$ are estimated in 
the case of $A\times N$ collisions being actually simulated. Note that the 
particle number $N$ is supposed to be quite a large number and the techniques 
of improvement in the speed are also used in both algorithms described by 
Mar\'{\i}n et al.,\cite{marin_1993}

\begin{itemize}

\item Cell-Crossing type (LCM + Cell-Crossing Event)

\begin{itemize}

\item The initial and last step ($\times 1$)
 
Linked-Cell Method --- $k_{LCM}\times N$
 
Computation of Event Time --- $k_{PP}\times 9 \times N_c\times N+k_{PC} 
\times 4\times N$
 
Construction of Complete Binary Tree --- $k_{CBT}\times N$
 
Update the final position of particles --- $k_{UPDATE} \times N$
 
\item Iteration Step (loop)

$(A\times N)\times (k_{PP}\times 9 \times N_c+k_{CBT} \times \log N) 
+ (B\times N)\times (k_{PC} \times 3+k_{CBT}\times \log N)$

\end{itemize}

where $N_c$ is the number of particles per sub-cell. 
The most important point is that the Cell-Crossing Event occurs at a certain 
ratio to Collision Events. Therefore, the additional computation of 
$B\times N$ times Cell-Crossing Events are needed when we want to simulate 
$A\times N$ times Collision Events. Since the complexity of the terms related 
to Cell-Crossing Events is always ${\cal O}(N)$, this is not negligible in the 
actual simulation.

\item EEPGM + DUTC

\begin{itemize}

\item Update of EEPGM ($\times C$)

EEPGM --- $k_{EEPGM}\times N$

Computation of Event Time --- $k_{PP}\times N_g \times N$

Search the Maximum Velocity --- $k_{MVS} \times N$
 
Construction of Complete Binary Tree --- $k_{CBT}\times N$

Update the final position of particles --- $k_{UPDATE}\times N$,

\item Iteration Step (loop)

$(A\times N)\times (k_{PP}\times N_g+k_{CBT}\times \log N) $

\end{itemize}

where $N_g$ is the averaged particle number of the mask.
Actually the value of $C$ is an order $C \sim A$, which is the same order of 
the frequency of updating EEPGM.

\end{itemize}

\noindent
Now, the comparison with order $N$ in both algorithm with constant coefficient 
is as follows:

\begin{itemize}

\item Cell-Crossing type

$(k_{LCM}+k_{PP}\times 9\times N_c\times (A+1) + k_{PC}\times (3\times B+1) 
+ k_{CBT} + k_{UPDATE})
\times N +(A\times k_{CBT} + B\times k_{CBT}) \times N \times \log N$

\item EEPGM + DUTC

$(k_{EEPGM}+k_{PP}\times 2\times N_g + k_{MVS} + k_{CBT} + k_{UPDATE})\times 
C \times N + (A\times k_{CBT}) \times N \times \log N$

\end{itemize}

\noindent
The most striking point is that the complexity of Cell-Crossing Events is of 
order of ${\cal O}(N \log N)$ (i.e., $B\times k_{CBT} \times N \times \log N$).
This result of analysis suggest that the efficiency of EEPGM + DUTC is better 
than for Cell-Crossing when the simulation with an enormous number of particle 
is performed. On the other hand, in the comparatively small particle system, 
the coefficient of $C\times N$ terms in EEPGM + DUTC may be larger than 
Cell-Crossing type.
However, the difference might be quite small, and it is impossible to estimate 
an exact coefficient of algorithms analytically. To the author's knowledge, 
the coefficient of $N$ terms is strongly dependent on the ratio of $N_c$ to 
$N_g$, and the rough estimation shows both algorithms are same when 
$N_g/N_c \sim 4$. Actually, the increase in computing the Cell-Crossing Events has 
almost no effect to the CPU time as the simulation is actually performed
in a system with a very large number particles.

\section{Empirical Evaluation}
\label{sec:5-0}
\noindent
When actually running the simulation, the order of the complexity 
is not so reliable, because it depends strongly on a constant coefficient when 
the number of the particle is relatively small. Moreover, the efficiency of 
the code changes significantly by the ability of a computer, the language, 
the performance of compiler, and the ability of programmers. Though a perfect 
comparison of efficiency of codes developed by the past workers is impossible, 
some publications showed how many particle collisions per CPU-hour can their 
codes be computed by their computers.
 
Mar\'{\i}n et al.\cite{marin_1993} simulated with hard disk 
system, and they also compared their code with the 
codes based on two main high-speed algorithms proposed by Rapaport
\cite{rapaport_1980} and Lubachevsky,\cite{lubachevsky_1991} respectively.
As a result, it was shown that the efficiency of the code of 
Mar\'{\i}n et al. was higher for the entire density region.
Therefore, the code proposed here should be compared with the code of 
Mar\'{\i}n et al., and is computed with equal number of 
particles ($N=2,500$). Mar\'{\i}n et al. achieved the maximum 
performance of 16.07 millions of collisions per one CPU hour on 
a SUN690 workstation. Note that only the highest performance is shown 
because it is different for different densities of 
the systems. On the other hand, the code proposed here achieved a maximum performance 
of 460 million collisions per one CPU hour (Alpha600 compatible, DEC-Fortran).
It was found that a high efficiency was realized even if
the performance of the machine was reduced.
 
Note that the workstation of our laboratory could actually simulate a
2,500,000 particle system. In this simulation, the amount of installed 
memory was $250$ megabytes and the computation performance was 210 million
collisions per one CPU hour.

\section{Extension to Infinite Volume System}
\label{sec:6-0}
\noindent
In this section, a simple example in the open boundary system that does not 
have a ceiling is considered. This is the case when there is an energy source 
at the bottom of the system under uniform gravity.
 
The system is also divided into grids by EEPGM.
However, because the top of the system is open, the grid goes to the top of 
the system infinitely. This means that the number of arrays for the grid becomes 
infinity, and the simulation is impossible from finite memory. To overcome 
this difficulty, the hashing method, which is well known as the fastest 
searching algorithm, is applied to keep the number of arrays in finite size 
and to simulate the dynamics of the system with high-efficiency.
 
\begin{itemize}
\item Construction of Data Structure
 
Firstly, the serial number of grid $N_G (i_g,j_g)$ is defined by
 
\begin{equation}
N_G=N_{gx}(j_g-1)+i_g,
\end{equation}
 
where $N_{gx}$ and $i_g (1 \leq i_g \leq N_{gx})$ are the total number of 
grids and the index of grid in the horizontal direction, respectively; 
$j_g (1 \leq j_g \leq \infty)$ refers to the index of grid in the vertical direction.
In addition, the maximum number of the serial grid $N_{Gmax}$ is calculated 
by $N_{Gmax}=N_{gx}(j_{gmax}-1)+i_{gmax}$; the pair $(i_{gmax}, j_{gmax})$ is 
at the maximum grid pair of containing the particle.
 
The serial grid $N_G (1 \leq N_G \leq N_{Gmax})$ is a one-dimensional array, 
in which particle number or $0$ is listed. This is called the one-dimensional 
{\it Virtual Array}. When $j_{max}$ is large value, there are many $0$'s 
in the one-dimensional {\it Virtual Array}. Though there is only information 
that a particle is not just in grid, this $0$ relates to the memory capacity 
directly. Now {\it Virtual Array} is compressed. After all, the only 
information necessary is the particle number and its index of grid. Therefore, 
one-dimensional integer arrays are prepared for $A(N), BX(N), BY(N)$, and 
grids of $0$ in {\it Virtual Array} are ignored, and then packed in order 
from the end; $A(N)$ stores the particle numbers only, and $BX(N)$ and $BY(N)$ 
stores indexes of $i_g$ and $j_g$ for each particle, respectively.
If you want to know whether there is a particle in grid $(i'_g, j'_g)$, you 
need only search the index-pairs correspond to $(i'_g, j'_g)$ in the lists 
of $BX(N), BY(N)$ linearly. However, this process is inefficient because the 
complexity of ${\cal O}(N)$ is necessary in searching.

Therefore, the hashing method, otherwise known as an algorithm which can realize the 
searching with  ${\cal O}(1)$ is applied. The following simplest hashing 
function is explained here as an example though various hashing functions can 
be considered and there is still room for development.

First, a hashing function is defined by

\begin{equation}
k=INT(\frac{N_G-1}{L})+1
\label{eqn:hash}
\end{equation}

which means that the {\it Virtual Array} is equally divided by the length $L$ 
(e.g., $5 \sim 10$) and key $k$ is calculated corresponding to serial number $N_G$.
Then, $k_{max}=INT(N_{Gmax}/L)$ is calculated using $N_{Gmax}$. Then additional 
arrays $C(k_{max}), D(k_{max})$ are prepared, these arrays are restored in $C(k)$ 
where $k$ begins in $A(N)$ and in $D(k)$ how big the arrays for each $k$.
In this case, necessary arrays are $A(N),BX(N),BY(N),C(k_{max}),D(k_{max})$, 
and these are confined to a finite value. Since additional arrays needed to use the 
hashing method are only $C(k_{max}), D(k_{max})\ (k_{max} < N$), the amount 
of memory used is only slightly increased in comparison with the linear searching.

\item Searching Process

In order to know what the particle number is in the grid ($i'_g,j'_g$), 
the following process is carried out.
First, a serial number is calculated by $N'_G=N_{gx}(j'_g-1)+i'_g$.
Next, $k'$ is calculated by the equation of hashing function (\ref{eqn:hash}).
Then, the searching ranges of $BX, BY$ are limited only to index of 
[$C(k') \sim C(k')+D(k')-1$] ($\leq L$). If there are equal pairs of grids
($i'_g,j'_g$) as a result of the searching of $BX, BY$, the index 
$s$ of $BX, BY$ reveals $A(s)$, which is the particle number in the grid 
($i'_g,j'_g$).
When equal pair is not found in $BX, BY$, there is no particle in the grid.

This procedure becomes possible in high-speed simulation.
The complexity becomes ${\cal O}(1)$ instead of the linear search 
${\cal O}(N)$, because a search is only carried out on the length of 
$L$ with hashing method.

\end{itemize}

A computation is possible for other boundary conditions with the same procedure 
if one-dimensional {\it Virtual Array} can be created. Therefore, high-speed 
simulation is possible in principle for any kind of boundary to be applied.
There is room for improvement in hashing functions, which is the easiest 
when dividing equally, because it is obviously inappropriate when particles are 
distributed heterogeneously.

The strategy of EEPGM has an advantage that it is easily extended.
The ease of a development is an important factor of the evaluation of an algorithm.

One problem in low-density systems is the overwhelming increase of the arrays 
assigned to grid in comparison to the number of the particles.
This is not desirable as the memory capacity is the same of as that of
infinite system. However, the arrays for the grid are compressed to the size of the 
particle number in the same way as described above. Since supplementary arrays 
are made by the hashing method and information on neighboring grids is efficiently 
obtained, there is no problem for both efficiency and memory capacity.

\section{Concluding remarks}
\label{sec:9-0}
\noindent
In this paper, we developed an algorithm for a hard disk system without using 
Cell-Crossing Event, which is simple, efficient and easy to 
extend.
EEPGM can be easy to extend because of its simplicity, which can never be 
realized in LCM. One example is the system that hard disks with various size of 
diameters coexist. Though there was a limitation in the degree of the 
poly-dispersion with EPGM described to Buchholtz and P\"oschel,
\cite{buchholtz_1993} EEPGM can be applied easily to those systems.
First, the grid is made based on the smallest particle radius in the system.
Next, we have only to check the nearest grids by using a suitable mask of
 the bigger level when the poly-dispersity increases.
This way, EEPGM has a wider application than EPGM.
 
This code achieved the fastest simulation speed in the world; about 460 million 
of collisions per CPU hour for the 2500 disk system on the VT-Alpha-600.
Since the order of complexity is ${\cal O} (\log N)$, the increase of complexity 
is slow when the particle number increases. Now, we can carry out large-scale 
molecular dynamics simulation $(\sim 10^6)$ on the usual Workstation in our 
laboratory. 
 
Hard particle simulation is a powerful tool for studying the fluidized state 
described by the kinetic theory or hydrodynamics.
Therefore, the large-scale simulation in the hard particle system will 
become an important subject.
 
Finally, the algorithm in this paper is suitable for the scalar machine, 
and the development of an algorithm for the parallel machine is the subject 
of a future study. 
Note that we discuss only hard disk systems in 2D for simplicity, but 
an extension to hard spheres system in 3D is easy to be carried out.

\nonumsection{Acknowledgements}
\noindent
The auther thanks Prof. Nakanishi for reading the manuscript and making a 
number of helpful suggestions. He also gratefully acknowledge helpful 
discussions with Dr. Muranaka on several points in the paper. A part of the 
computation in this work was done by the facilities of the Supercomputer 
Center, Institute for Solid State Physics, University of Tokyo.

\nonumsection{References}

\end{document}